\documentclass[PRL,twocolumn,groupedaddress,nofootinbib,superscriptaddress]{revtex4-2}
\usepackage{graphicx}
\usepackage{dcolumn}
\usepackage{bm}
\usepackage{times}
\usepackage{epstopdf}
\usepackage[version=3]{mhchem} 
\usepackage{xcolor}

\usepackage{mathrsfs}
\usepackage{lipsum}
\usepackage{amsmath}
\usepackage{amsfonts}
\usepackage{array}
\usepackage{color}
\usepackage{ulem}

\newcommand{\sibt}{Sn$_{\rm{1-x}}$In$_{\rm{x}}$Bi$_{\rm{2}}$Te$_{\rm{4}}$}
\newcommand{\sbt}{SnBi$_{\rm{2}}$Te$_{\rm{4}}$}

\begin{document}

\title{Superconductivity by alloying the topological insulator \sbt }

\author{Michael A. McGuire}
\email{mcguirema@ornl.gov  \\  \\ Notice: This manuscript has been authored by UT-Battelle, LLC under Contract No. DE-AC05-00OR22725 with the U.S. Department of Energy. The United States Government retains and the publisher, by accepting the article for publication, acknowledges that the United States Government retains a non-exclusive, paid-up, irrevocable, world-wide license to publish or reproduce the published form of this manuscript, or allow others to do so, for United States Government purposes. The Department of Energy will provide public access to these results of federally sponsored research in accordance with the DOE Public Access Plan (http://energy.gov/downloads/doe-public-access-plan). }
\author{Heda Zhang}
\affiliation{Materials Science and Technology Division, Oak Ridge National Laboratory, Oak Ridge, Tennessee 37831 USA}
\author{Andrew F. May}
\affiliation{Materials Science and Technology Division, Oak Ridge National Laboratory, Oak Ridge, Tennessee 37831 USA}
\author{Satoshi Okamoto}
\affiliation{Materials Science and Technology Division, Oak Ridge National Laboratory, Oak Ridge, Tennessee 37831 USA}
\author{Robert G. Moore}
\affiliation{Materials Science and Technology Division, Oak Ridge National Laboratory, Oak Ridge, Tennessee 37831 USA}
\author{Xiaoping Wang}
\affiliation{Neutron Scattering Division, Oak Ridge National Laboratory, Oak Ridge, Tennessee 37831, USA}\author{Cl\'ement Girod}
\affiliation{Los Alamos National Laboratory, Los Alamos, New Mexico, 87545, USA}
\author{Sean M. Thomas}
\affiliation{Los Alamos National Laboratory, Los Alamos, New Mexico, 87545, USA}
\author{Filip Ronning}
\affiliation{Los Alamos National Laboratory, Los Alamos, New Mexico, 87545, USA}
\author{Jiaqiang Yan}
\affiliation{Materials Science and Technology Division, Oak Ridge National Laboratory, Oak Ridge, Tennessee 37831 USA}

\begin{abstract}
Alloying indium into the topological insulator \sbt\ induces bulk superconductivity with critical temperatures $T_c$ up to 1.85\,K and upper critical fields up to about 14\,kOe. This is confirmed by electrical resistivity, heat capacity, and magnetic susceptibility measurements. The heat capacity shows a discontinuity at $T_c$ and temperature dependence below $T_c$ consistent with weak coupling BCS theory, and suggests a superconducting gap near 0.25\,meV. The superconductivity is type-II and the topological surface states have been verified by photoemission. A simple picture suggests analogies with the isostructural magnetic topological insulator \ce{MnBi2Te4}, in which a natural heterostructure hosts complementary properties on different sublattices, and motivates new interest in this large family of compounds. The existence of both topological surface states and superconductivity in \sibt\ identifies these materials as promising candidates for the study of topological superconductivity.
\end{abstract}

\maketitle


Designing new quantum materials, and functional materials in general, often requires combining multiple phenomena or behaviors into a single sample or system. This is true for topological materials, where combining topology with magnetism can generate Weyl semimetals and Chern and axion insulators  \cite{chang2013experimental, weng2015quantum, xiao2018realization, liu2019magnetic}, and combining topology with superconductivity can produce Majorana zero modes \cite{fu2008superconducting, fu2010odd, sato2017topological, frolov2020topological}. These combinations can be achieved in several ways, and are often realized in designer heterostructures where dissimilar materials are physically stacked or grown together as thin films or nanostructures \cite{lutchyn2010majorana, xiao2018realization, frolov2020topological}. In some cases, functionalities can be combined in a single material \cite{chang2013experimental, zhang2018observation}. One approach is to employ crystal structures that have separate sublattices that each host one of the desired behaviors. An excellent example of this is given by the combination of magnetism and topology in \ce{MnBi2Te4} and related materials \cite{li2019intrinsic, he2020mnbi2te4}. The crystal structures of these compounds contain ordered arrangements of magnetic MnTe layers and topological \ce{Bi2Te3} layers. This has led to intense study of these materials and reports of Chern and axion insulating behavior, and the quantum anomalous Hall effect \cite{deng2020quantum, liu2020robust, ge2020high, ovchinnikov2021intertwined, zhao2021even}.

Combining topology with superconductivity is expected to be possible in triplet or p-wave superconductors \cite{ivanov2001non, fu2010odd}, but the more common approach is based on the Fu-Kane model, in which the spin-momentum locked surface states in at topological insulator interact with the s-wave superconductivity in another material via proximity effects \cite{fu2008superconducting}. These two phenomena may arise from separate sources in a single material, giving rise to an intrinsic proximity effect, for example when a topological semimetal becomes superconducting. The iron-based superconductor FeSe$_{1-x}$Te$_x$ is an example of this, with topological surface states related to the strong spin orbit coupling of Te and superconductivity related primarily to Fe \cite{zhang2018observation, Wang2015Topological}

\begin{figure}
\begin{center}
\includegraphics[width=3.25in]{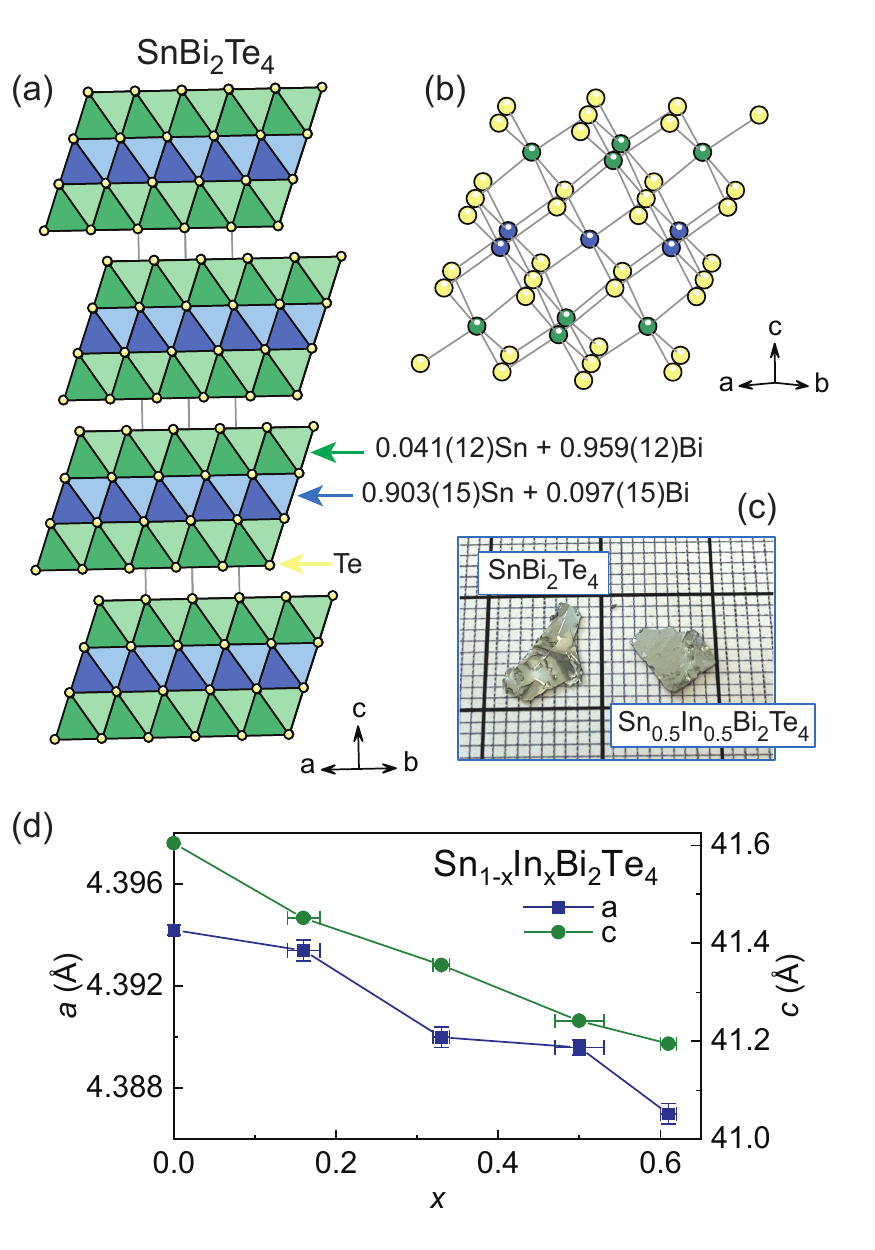}
\caption{\label{fig:structure}
Structure of \sibt. (a) A view of the stacking of the septuple layers with cation site occupancies for \sbt\ (x = 0) determined by neutron single crystal diffraction. (b) The internal structure of the septuple layers showing the octahedral coordination environments. (c) Crystals with x\,=\,0 and 0.5 on a mm/cm grid. (d) The x-dependence of the lattice parameters determined by powder x-ray diffraction at room temperature.
}
\end{center}
\end{figure}

Here we explore this idea using the same materials family to which \ce{MnBi2Te4} belongs. This family, typified by the \ce{GeAs2Te4} structure type, contains \ce{AM2X4} with A\,=\,Mn, Ge, Sn, Pb, M\,=\,As, Sb, Bi, and X = Se, Te. The structure of \sbt\ is shown in Fig. \ref{fig:structure}a. Since these compounds are formally charge balanced and contain many heavy main group elements, they form small band gap semiconductors with strong spin-orbit coupling. Thus, they have attracted attention for their thermoelectric performance \cite{kuznetsova2000thermoelectric, shelimova2004crystal, zhang2010electronic} and more recently for their electronic topology. Electronic structure calculations show that this family contains many topologically nontrivial compounds \cite{jin2011candidates, eremeev2012atom, menshchikova2011ternary, vergniory2015electronic, wang2022highly}.
Topological surface states have been observed experimentally in \ce{GeSb2Te4} \cite{nurmamat2020topologically}, \ce{GeBi2Te4} \cite{neupane2012topological, okamoto2012observation, li2021topological}, \ce{SnBi2Te4} \cite{li2021topological, fragkos2021topological}, \ce{SnSb2Te4} \cite{niesner2014bulk}, and \ce{PbBi2Te4} \cite{kuroda2012experimental}.  The present work concerns specifically \sbt\ \cite{zhukova1972crystal, kuropatwa2012thermoelectric}, which has been identified as a 3D topological insulator based on parity of the valence band states at the $\Gamma$ point and the observation of topological surface states with angle resolved photoemission spectroscopy (ARPES) measurements \cite{menshchikova2011ternary, eremeev2012atom, vergniory2015electronic, fragkos2021topological, li2021topological}. The electronic structure calculations and ARPES data suggest \sbt\ is a small band gap semiconductor or semimetal, with Dirac surface states overlapping the bulk valance band.

To introduce superconductivity into this family, a superconducting compound of composition (Ge,Sn,Pb)(Se,Te) is desired (in analogy of introducing magnetism based on the magnetic compound MnTe). Fortunately, there are reports of tuning some of these binary compounds into superconducting states by alloying. This includes Pb$_{1-x}$Tl$_x$Te ($T_c \approx 1.5$\,K) \cite{matsushita2006type}, Pb$_{1-x}$In$_x$Te ($T_c \approx 4.8$\,K) \cite{smylie2022full}, and Sn$_{1-x}$In$_x$Te ($T_c \approx 4.7$\,K) \cite{balakrishnan2013superconducting, zhong2013optimizing}.  Here we employ indium doping to tune the the electronic ground state in the topological insulator \sbt.

In this Letter, we show that indium doping does induce superconductivity in \sibt. The superconductivity is bulk and robust in the single crystals grown for this study. This is evidenced by sharp transitions to zero resistance, large volume fractions inferred from strong diamagnetism, and sharp and large specific heat capacity anomalies. The heat capacity shows behavior consistent with weak coupling BCS theory and a single superconducting gap. For the highest indium content achieved here ($x = 0.61$), $T_c$ reaches 1.85\,K, with a maximum upper critical field near 14\,kOe. The dependence of $T_c$ on $x$ suggests higher transition temperatures could be realized at higher indium concentrations. The expected topological surface states are confirmed by ARPES measurements for $x = 0.5$, in analogy with the parent phase \sbt. Thus, this work establishes \sibt\ as an interesting candidate topological superconductor, and motivates detailed spectroscopic studies to investigate the coupling between superconductivity and topology and the potential presence of Majorana zero modes.


Experimental details of the growth and chemical/structural characterization of the crystals and measurements of their physical properties are included in the Supplemental Material \cite{supp}. The crystals' compositions determined by energy dispersive spectroscopy (EDS) and their lattice parameters refined from powder x-ray diffraction data are shown in Table \ref{Table}. The EDS results are normalized to give a total of seven atoms in the formula unit, and are consistent with the formula \textit{M}\ce{Bi2Te4} with \textit{M}\,=\,Sn, In. In the following, the samples will be referred to by the value of \textit{x} in the chemical formula \sibt\ determined by EDS (Table \ref{Table}).

\begin{table*}
\caption{\sibt\ crystal compositions, room temperature lattice parameters (a, c), normal state Sommerfeld coefficients ($\gamma$), superconducting transition temperatures ($T_c$), specific heat discontinuities at $T_c$ ($\Delta c_P$), and upper critical field at $T = 0$ estimated in the WHH approximation for fields along the c-axis ($H_{c2}(0)^{WHH}$).  For $x \geq 0.33$ the average and the standard deviation calculated from the three values determined from resistivity, magnetic susceptibility, and heat capacity data are listed. }
\begin{tabular}{lccccccc}
\hline															
EDS composition	&	$x$ in 	&	a 	&	c	&	$\gamma$ 	&	$T_c$	&	$\Delta c_P (\gamma T_c)^{-1}$	&	$H_{c2}(0)^{WHH}$ 	\\
Sn:In:Bi:Te	&	\sibt	&	({\AA})	&	 ({\AA})	&	(J mol-FU$^{-1}$ K$^{-2}$)	&	(K)	&		&	(kOe)	\\
\hline															
0.99(1):0:1.96(2):4.05(2)	&	0	&	4.3942(2)	&	41.604(1)	&	0.0027(1)	&	--	&	--	&	--	\\
0.83(3):0.15(2):1.99(3):4.03(3)	&	0.16	&	4.3934(4)	&	41.452(3)	&	--	&	0.43	&	--	&	--	\\
0.68(1):0.34(1):1.97(2):4.02(2)	&	0.33	&	4.3900(4)	&	41.356(2)	&	0.0055(2)	&	1.22(5)	&	1.23	&	10.3	\\
0.47(3):0.47(2):2.00(1):4.06(3)	&	0.5	&	4.3896(3)	&	41.242(2)	&	0.0065(1)	&	1.65(4)	&	1.45	&	13.2	\\
0.38(1):0.59(2):2.00(1):4.03(2)	&	0.61	&	4.3870(4)	&	41.195(3)	&	0.0065(4)	&	1.85(5)	&	1.51	&	13.8	\\
\hline															
\end{tabular}\
\label{Table}
\end{table*}

%
%


%
\begin{figure}
\begin{center}
\includegraphics[width=3.25in]{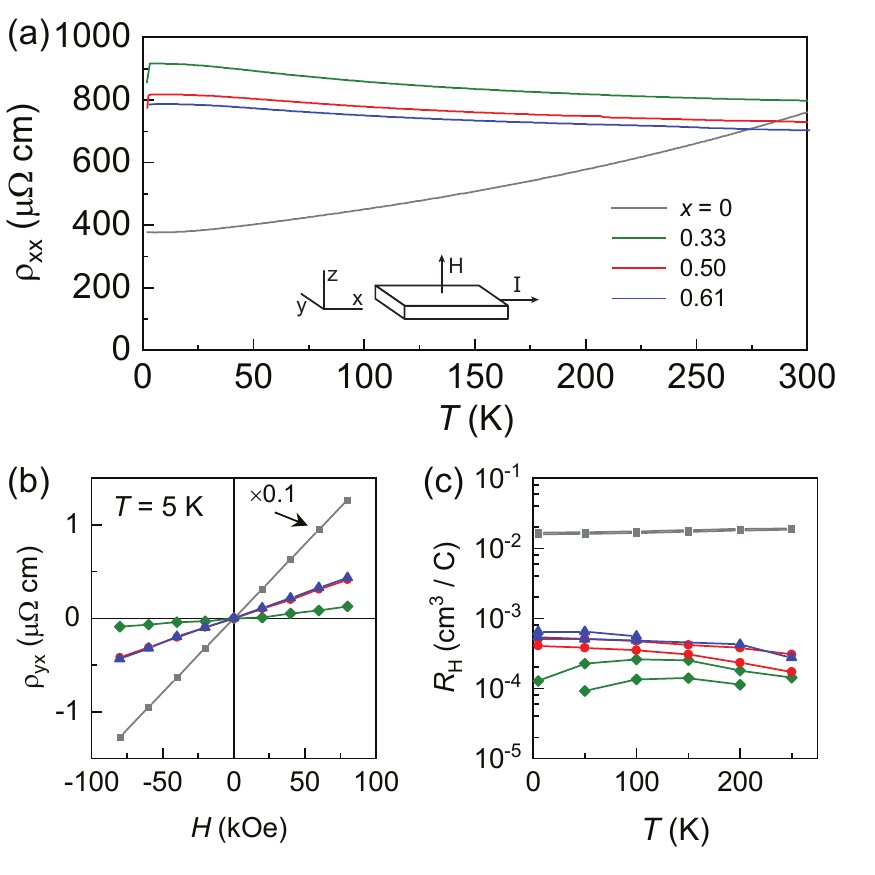}
\caption{\label{fig:transport}
Normal state electrical transport in \sibt. (a) Resistivity measured in zero magnetic field. (b) Transverse resistivity measured in fields from -80 to 80\,kOe. The data has been antisymmetrized to exclude contributions from longitudinal voltages. (c) Hall coefficient determined from the slope of $\rho_{yx}(H)$. Panel (c) shows data from two samples of each composition.
}
\end{center}
\end{figure}

All of the powder x-ray diffraction data from ground crystals of \sibt\ taken from the growths discussed here were consistent with the \ce{GeAs2Te4} structure type (see Supplemental Material \cite{supp}). The structure was further confirmed for \ce{SnBi2Te4} using single crystal neutron diffraction at room temperature. The van der Waals layered structure is shown in Figure \ref{fig:structure}a and \ref{fig:structure}b. The neutron diffraction was used to refine the degree of cation mixing, known to be common in this family of materials \cite{zou2018atomic, liu2021site, zhang2022cation}. The refined Sn and Bi concentrations on their two crystallographic sites is shown in Figure \ref{fig:structure}a. Substitution of In into \ce{SnBi2Te4} is seen to compress the lattice both in the plane and along the stacking direction (Figure \ref{fig:structure}d). Based on the measured stoichiometries (Table \ref{Table}), it is clear that In primarily replaces Sn as expected, and the decrease in lattice parameters with increasing In concentration is consistent with the reported behavior in Sn$_{1-x}$In$_x$Te \cite{zhong2013optimizing}. Subtleties of how In may be distributed along with Sn and Bi on the two cation sites (see Fig. \ref{fig:structure}a for \sbt) are beyond the scope of this work, and the simple formula \sibt\ suffices to describe the materials for the present purposes.

Results of resistivity and Hall effect measurements in the normal state are shown in Figure \ref{fig:transport}. \sbt\ behaves like a poor metal, heavily doped semiconductor, or semimetal, with a resistivity changing by about a factor of two between room temperature and 2\,K. Indium substitution has only a small effect on the value of the resistivity at room temperature, but changes the temperature dependence to show slight increases upon cooling. Similar behavior was observed when In was substituted into SnTe \cite{balakrishnan2013superconducting}, although in that case the temperature dependence remained metal-like and the change in resistivity could be attributed to defect scattering increasing the residual resistivity.

In \sibt\ transverse resistivities $\rho_{yx}$ changed linearly with applied field at all temperatures, and are shown Figure \ref{fig:transport}b for 5\,K. The Hall coefficients determined from the slopes of $\rho_{yx}$ vs H are shown in Figure \ref{fig:transport}c. Positive values indicate conduction is dominated by holes in all samples. Comparing \sbt\ to the indium containing samples shows that the substitution significantly reduces $R_H$. In a single band model this would indicate In dopes holes into the system, consistent with the relative positions of Sn and In on the periodic table. However, no clear trend is seen in $R_H$ among the In containing samples. If two carrier types are present or multiple carrier pockets (Ref. \citenum{fragkos2021topological}) become involved in the transport, the interpretation of the change in $R_H$ with $x$ is not straightforward. 
Nevertheless, the data does suggest that indium introduces holes into the system, as expected. This is supported by Seebeck coefficient measurements on crystals with $x = 0$ and $x = 0.61$ as well (see \cite{supp}). The Seebeck coefficient for \sbt measured at 300\,K is 76\,$\mu$V/K, indicating p-type behavior and in agreement previous reports \cite{kuropatwa2012thermoelectric}. Indium substitution decreases this value, to 23\,$\mu$V/K for $x = 0.61$.

\begin{figure}
\begin{center}
\includegraphics[width=3.25in]{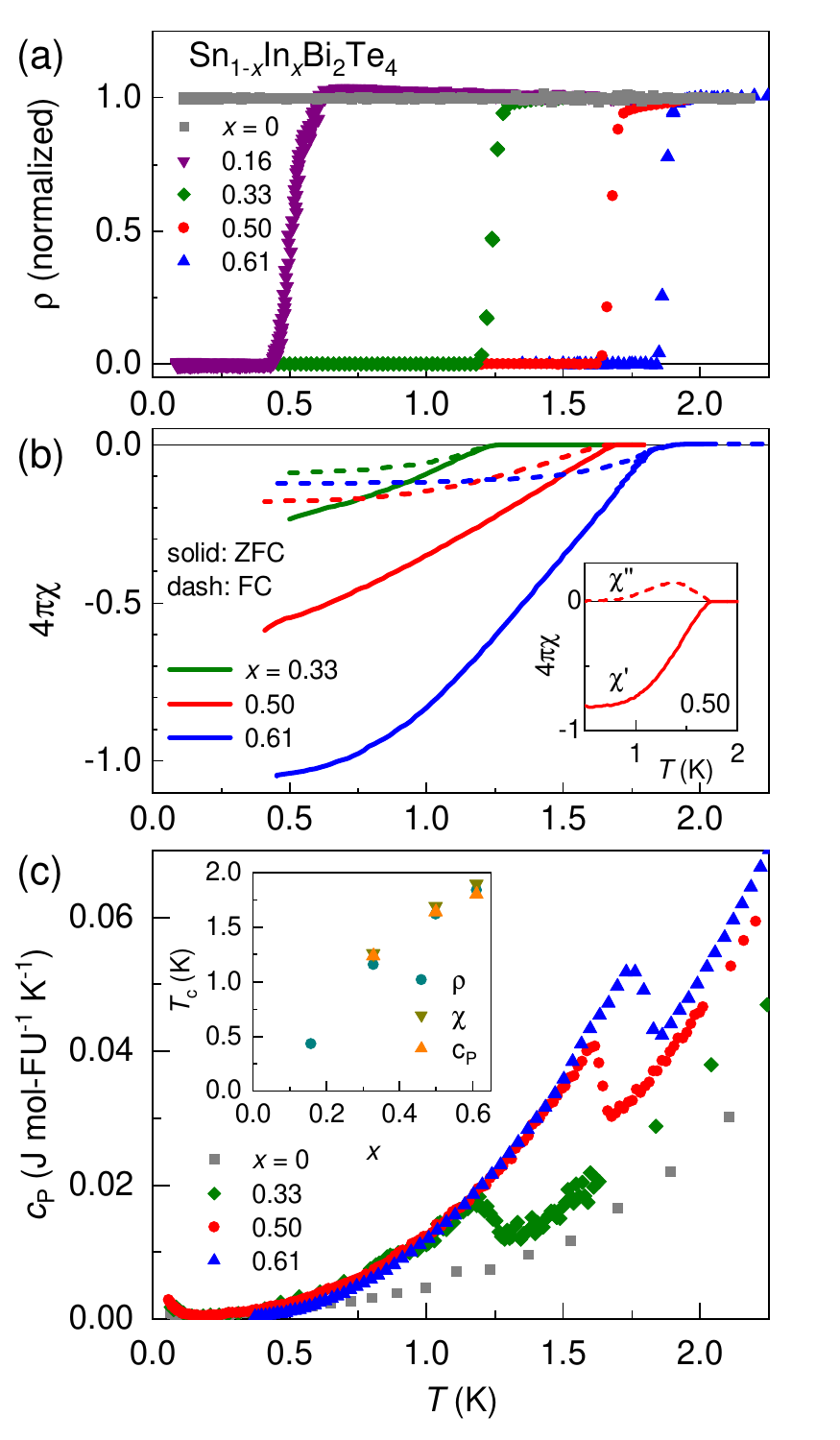}
\caption{\label{fig:SCsummary}
Evidence for bulk superconductivity in \sibt. (a) Resistivity, normalized in the normal state for comparison. (b) dc magnetic susceptibility, measured on warming in a 5\,Oe in-plane field after zero-field-cooling (ZFC) and field-cooling (FC), with ac susceptibility results shown in the inset. (c) Specific heat capacity per formula unit (FU). The inset in (c) shows the values of $T_c$ determined by the temperatures at which $\rho$ reaches zero, the onsets of diamagnetism, and the midpoints of the heat capacity jumps.
}
\end{center}
\end{figure}

The superconducting transitions in \sibt\ are demonstrated in Figure \ref{fig:SCsummary}. This figure shows results of electrical resistance, magnetization, and heat capacity measurements. Figure \ref{fig:SCsummary}a shows sharp transitions to zero resistance states for $x \geq 0.16$, and we define $T_c$ as the temperature at which the resistance reaches zero. The dc magnetic susceptibility data shown in the main panel of Figure \ref{fig:SCsummary}b demonstrate large superconducting volume fractions, and show divergence between zero-field-cooled and field-cooled data. The latter is an indication of type-II superconductivity and the presence of a mixed state when cooling through $T_c$ in an applied field. We define the transition temperature from this data as the onset of diamagnetism. The inset shows results of ac magnetic susceptibility for a crystal with $x = 0.50$. The specific heat capacity data in Figure \ref{fig:SCsummary}c reveals sharp $\lambda$-anomalies at the superconducting phase transition, and we define $T_c$ as the midpoint of the specific heat capacity jump.

The inset in Fig. \ref{fig:SCsummary}c shows the $T_c$ values determined from these measurements. Good agreement is seen among the values determined from three types of measurements. The transition temperature increases with $x$ up to the maximum indium content achieved in this study ($x = 0.61)$. Table \ref{Table} contains the $T_c$ values. For $x \geq 0.33$ the average and the standard deviation calculated from the three values determined from resistivity, magnetic susceptibility, and heat capacity data are listed. It is likely that the dependence of $T_c$ on $x$ in \sibt\ is primarily related to changes in the chemical potential associated with In substitution. As noted above, the introduction of In does compress the lattice (Fig. \ref{fig:structure}d), but the  roles played by electronic tuning and chemical pressure in determining the ground state cannot be distinguished reliably using the present data.

\begin{figure}
\begin{center}
\includegraphics[width=3.25in]{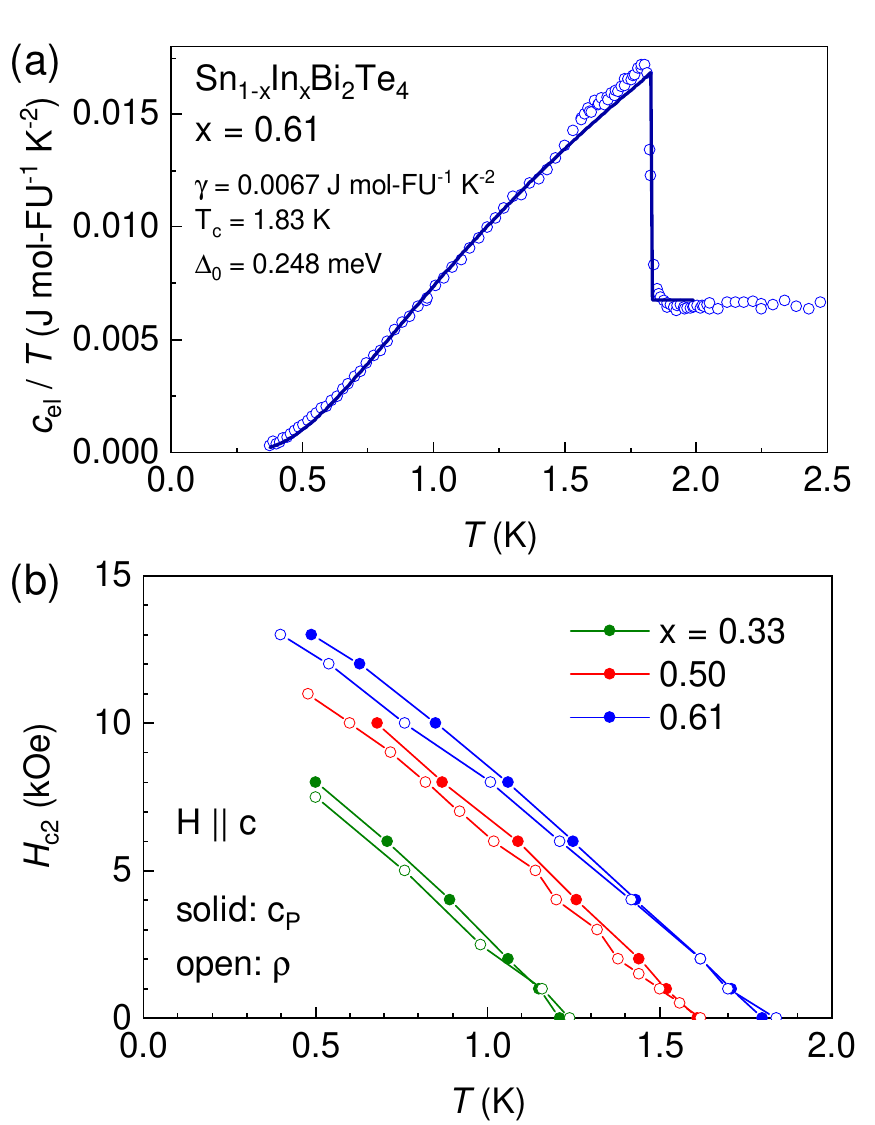}
\caption{\label{fig:hc-PD}
(a) A fit of the electronic heat capacity near and below $T_c$ for \sibt\ with $x = 0.61$. See text for details. (b) Superconducting phase diagrams for \sibt\ determined from $c_P$ and $\rho$ measurements performed in magnetic fields directed out of the plane (parallel to the c-axis). Data used to construct the plots are shown in \cite{supp}.
}
\end{center}
\end{figure}

Values of $\gamma$ determined from the fit of the normal state heat capacity (see \cite{supp}) are collected in Table \ref{Table}, and show a general increase with increasing $x$. This is consistent with indium substitution increasing the carrier concentration in \sibt. Weak coupling BCS theory predicts a specific heat discontinuity ($\Delta c_P$) at $T_c$ of 1.43$\gamma T_c$. The measured ratios $\Delta c_P/(\gamma T_c)$ are shown in Table \ref{Table}, and are close to the BCS weak coupling value.

Specific heat capacity data for the highest $T_c$ material ($x = 0.61$) was modeled assuming BCS behavior

\begin{eqnarray*}
C_\mathrm{BCS}=t\frac{d}{dt}\int_0^{\infty}\,dy(-\frac{6\gamma\Delta_0}{k_\mathrm{B}\pi^2})[f\ln{f}
+ (1-f)\ln{(1-f)}].
\end{eqnarray*}
where $t$\,=\,$T$/$T_\mathrm{c}$,
$f$\,=\,1/[$\exp$($E$/$k_\mathrm{B}$$T$)+1],
$E$\,=\,($\epsilon^2$+$\Delta^2$)$^{1/2}$,
$y$\,=\,$\epsilon$/$\Delta_0$, and $\Delta(T)/\Delta_0$ is the BCS temperature dependence of the gap taken
from the tables of M\"{u}hlschlegel \cite{Muhlschlegel}.

The data are well fit using a single superconducting gap $\Delta_0$ along with $\gamma$ and $T_c$ as three free parameters. The fitted curve and parameters are shown in Fig. \ref{fig:hc-PD}a. The fitted $T_c$ and $\gamma$ are consistent with the other determinations presented here (Table \ref{Table}). The fitted gap is 0.248\,meV, or 1.80\,$k_BT_c$. This is very close to the BCS weak-coupling value of $\Delta/(k_BT_c)$\,=\,1.76. Double integration of the fitted specific heat curve from 0 to $T_c$ gives the condensation energy of 4.8 mJ/mol-FU. The condensation energy is equal to $H_{c}^{2}/8\pi$ from which we obtain a thermodynamic critical field of 94 Oe.

To understand the superconducting phase diagram of \sibt, the effects of magnetic fields applied parallel to the c-axis were investigated using electrical resistivity and heat capacity measurements for $x = 0.33, 0.50, 0.61$ (data are shown \cite{supp}). The magnetic field suppresses superconductivity as expected. A plot of applied field vs $T_c$ determined from these measurements traces the upper critical field $H_{c2}$. These curves are shown in Figure \ref{fig:hc-PD}b. The Werthamer-Helfand-Hohenberg (WHH) theory gives an estimate of critical field at zero temperature based on the slope near $T_c$ in low fields: $H_{c2}(0) = -0.69T_c(dH_{c2}/dT)|_{T = T_c}$. Table \ref{Table} contains the results obtained by applying this to the phase diagram data derived from specific heat measurements. The $H_{c2}(0)$ values determined in this way are consistent with the behaviors seen down to the lowest temperatures studied (Figure \ref{fig:hc-PD}b). For $x = 0.61$ we find $H_{c2}(0) = 13.8$ kOe, which compared with the thermodynamic critical field of 94 Oe obtained from our heat capacity fit firmly establishes this compound as a type II superconductor. The broadening of the superconducting transition in response to the applied field (see \cite{supp}) supports this conclusion.

\begin{figure}
\begin{center}
\includegraphics[width=3.25in]{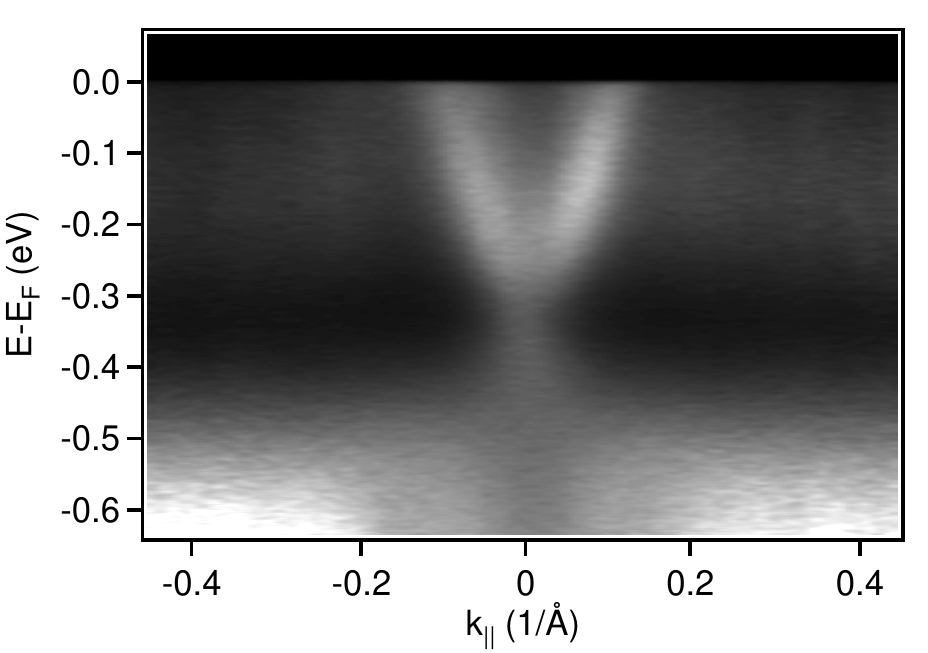}
\caption{\label{fig:arpes}
ARPES data for \sibt\ with $x = 0.5$ at the Brillouin zone center showing the linear dispersion from the topological Dirac surface states. Data were measured at $T \sim 8$\,K.
}
\end{center}
\end{figure}

To verify the topological nature of the doped \sibt\ family, ARPES data were taken at the Brillouin zone center ($\Gamma$) for $x = 0.5$ as described in the Supplemntal Material \cite{supp}.   Previous reports reveal Dirac surface states for SnBi$_2$Te$_4$ establishing the material as a topological insulator \cite{li2021topological}.  While the calculated band structure for SnBi$_2$Te$_4$ predicts a topological insulating band structure with the Dirac point located at $E-E_F \sim 0.1$ eV, the measured bands structure showed the Dirac point to be located at the Fermi level \cite{li2021topological, eremeev2012atom}.  For \sibt\ with $x = 0.5$ we observe the linearly dispersing topological surface states with the Dirac point located at $E-E_F = -0.35$ eV as shown in Fig. \ref{fig:arpes}.  The measured bands structure appears as a rigid shift of the electronic structure due to the doping and shows the topological nature of the family is not altered.


In summary, we have identified \sibt\ as an interesting cleavable material for the study of the interplay of non-trivial electronic topology and superconductivity. \sbt\ is known to be a topological insulator \cite{menshchikova2011ternary, eremeev2012atom, vergniory2015electronic, fragkos2021topological, li2021topological}. Here we show that indium substitution induces bulk superconductivity in \sibt\ while maintaining the presence of topological surface states near the Fermi level. For the highest indium concentrations, $T_c$ and $H_{c2}(0)$ reach approximately 1.85\,K and 14\,kOe, respectively. The specific heat capacity follows BCS behavior. The superconductivity is type-II, which could support vortices hosting Majorana zero modes in the Fu-Kane model \cite{fu2008superconducting}. The next steps are to experimentally and theoretically understand in detail how the topological surface states evolve with In content, how they interact with the superconductivity, and its resulting topological nature.
While this work as been considered primarily within the context of the Fu-Kane model \cite{fu2008superconducting}, the centrosymmetric structure and strong spin-orbit coupling make \sibt\ potentially compatible with the Fu-Berg model for bulk topological superconductivity as well, as proposed for Cu-doped \ce{Bi2Te3} \cite{fu2010odd}. Importantly, \sbt\ is a member of a large family of similar natural heterostructures, and as noted above, many are expected to have nontrivial topology. This suggests study of not only \ce{AM2X4} (or AX$\cdot$\ce{M2X3}) with A\,=\,Ge, Sn, Pb, M\,=\,As, Sb, Bi, and X = Se, Te, but also $n$(AX)$\cdot$\ce{M2X3} with thicker AX slabs, AX$\cdot m$(\ce{M2X3}) with additional \ce{M2X3} layers, and, in principle, $n$(AX)$\cdot m$(\ce{M2X3}). Clearly there is much interesting work to be done and we hope the present study motivates further research exploring \sibt\ and related compounds as a new family of candidate topological superconductors.

\section*{Acknowledgements}
This material is based upon work supported by the U.S. Department of Energy, Office of Science, National Quantum Information Science Research Centers, Quantum Science Center. AFM acknowledges support from the U.S. Department of Energy, Office of Science, Basic Energy Sciences, Materials Sciences and Engineering Division (heat capacity measurements). A portion of this research used resources at the Spallation Neutron Source, a DOE Office of Science User Facility operated by the Oak Ridge National Laboratory.


%


\clearpage
\newpage

\setcounter{equation}{0}
\setcounter{figure}{0}
\setcounter{table}{0}
\setcounter{page}{1}
\makeatletter
\renewcommand{\theequation}{S\arabic{equation}}
\renewcommand{\thefigure}{S\arabic{figure}}
\renewcommand{\thetable}{S\arabic{table}}
\renewcommand{\bibnumfmt}[1]{[S#1]}
\renewcommand{\citenumfont}[1]{S#1}

\section{Supplementary Materials}

Single crystals of \sbt\ were grown out of Te flux. Sn shot, Bi pieces, and Te shot in the ratio of 1(\sbt):4Te were placed in a 2ml alumina growth crucible of a Canfield crucible set. The crucible set was then sealed under vacuum inside of a silica ampoule. The sealed ampoule was heated to 900$^{\circ}$C to obtain a homogeneous melt and then cooled over 10 days to 480$^{\circ}$C. At this temperature, Te flux was decanted from the \sbt\ single crystals.

Single crystals of \sibt\ were grown by heating a stoichiometric mixture of the starting elements at 565$^{\circ}$C for over 2 weeks. The melting behavior of \sibt\ was not found in the literature. Our test growths suggest that (1) In-rich \sibt\ melts, most likely peritectically, around 570$^{\circ}$C, and (2) the appropriate temperature window suitable for crystal growth is narrow, similar to  MnBi$_2$Te$_4$ \cite{yan2019crystal}. The In content $x$ in the \sibt\ crystals grown for this study was limited to about 0.61. Attempts to increase this led to the formation of secondary phases coexisting with Sn$_{\sim 0.4}$In$_{\sim 0.6}$Bi$_2$Te$_4$.  This suggests that the solubility limit of In in \sbt\ is near this value under the range of conditions used here. Increasing this value, or realizing \ce{GeAs2Te4}-type \ce{InBi2Te4}, may require different approaches (high pressure or significantly altered melt compositions for example).

\begin{figure*}
\begin{center}
\includegraphics[width=3.5in]{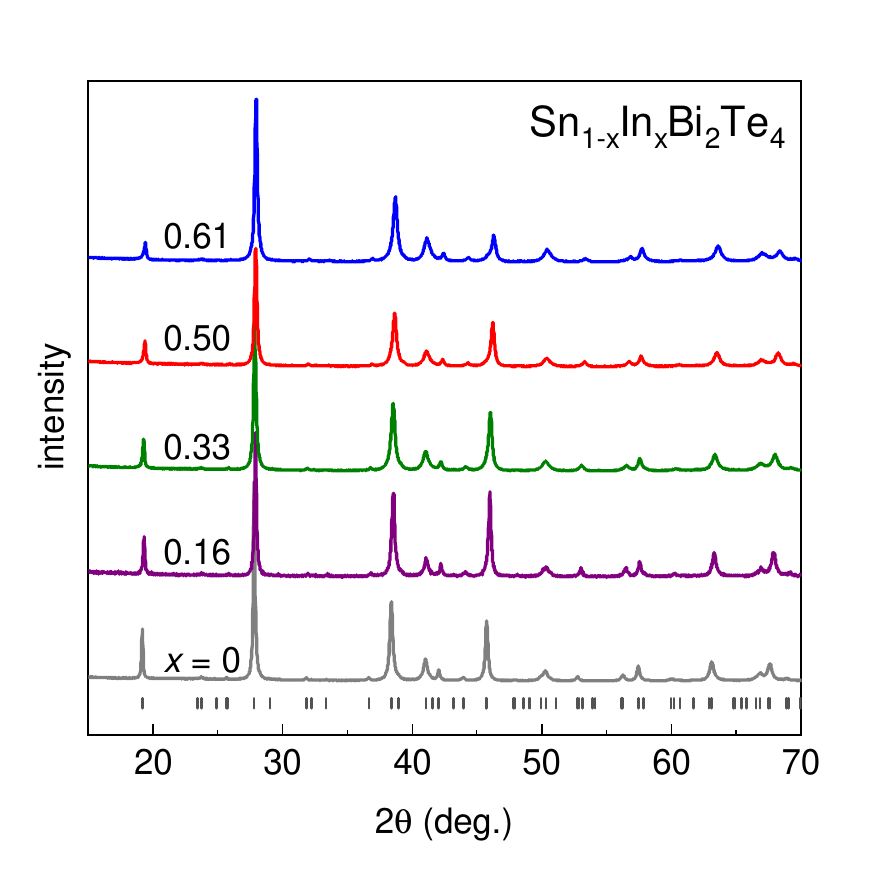}
\caption{\label{fig:SuppFig-pxrd}
Powder x-ray diffraction data from \sibt\ collected using Cu K$_{\alpha 1}$ radiation. The patterns are offset vertically for clarity. The tick marks below the $x = 0$ data locate reflections from the \ce{GeAs2Te4} structure type that has been previously reported for \sbt\ \cite{kuropatwa2012thermoelectric}.
}
\end{center}
\end{figure*}

Powder X-ray diffraction using a PANalytical X'Pert Pro MPD diffractometer was used to confirm the single phase nature of the crystals and determine their lattice parameters at room temperature. Data collected at room temperature is shown in Fig. \ref{fig:SuppFig-pxrd}. The samples were prepared by grinding single crystals.

The compositions of the crystals were measured using energy dispersive spectroscopy (EDS) with a Hitachi TM3000 scanning electron microscope and Bruker Quantax70 spectrometer. Data from at least three spots on at least three different crystals were used to calculate the average compositions.

Measurements of thermodynamic and electronic properties were performed using commercial cryostats from Quantum Design (Dynacool, PPMS, MPMS-3) with the addition of an adiabatic demagnetization refrigerator option for transport measurements, $^{\rm{3}}$He options for transport, heat capacity, and ac and dc magnetization measurements, and a dilution refrigerator option for heat capacity measurements. Contacts for transport measurements were made using silver paste (Dupont 4929N).

\begin{figure*}
\begin{center}
\includegraphics[width=3.5in]{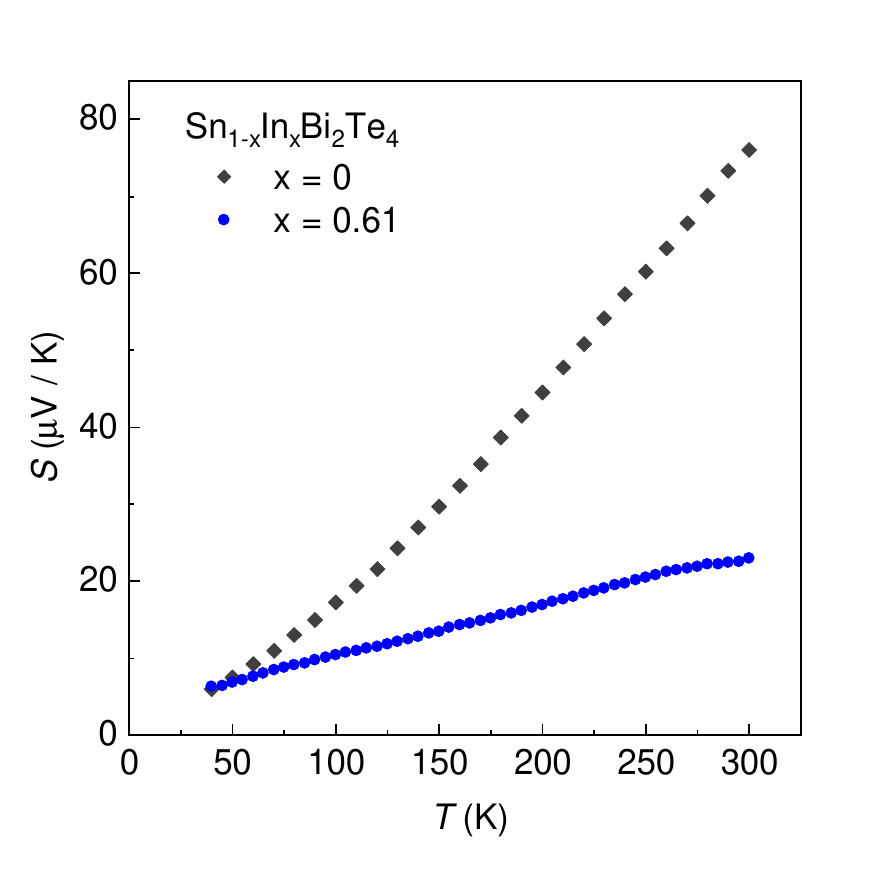}
\caption{\label{fig:SuppFig-Seebeck}
Seebeck coefficient data measured for \sibt\ with $x = 0, 0.61$.
}
\end{center}
\end{figure*}

Seebeck measurements were performed on \sibt, x = 0, 0.61, in the temperature range between 300 K to 40 K under zero magnetic field. At each temperature, a heater attached to the sample was turned on and the temperature at the hot end ($T_{hot}$) and cold end ($T_{cold}$) was monitored with thermocouples until they both became stable within 20\,mK. $T_{hot}$ and $T_{cold}$ and the Seebeck voltage (V) were recorded then the heater was turned off. After waiting for a time equal to 1.2 times of the heating period the equilibrium values of $T_{hot}$, $T_{cold}$ and V were recorded. The Seebeck coefficient is obtained by $S =[(V_{on}-V_{off}) /l_v]/[(\Delta T_{on} - \Delta T_{off} )/l_t]$, where $\Delta T$ is the temperature difference between hot end and cold end, on/off denote the equilibrium state when heater was on/off, and $l_v$ and $l_t$ denote the distance between voltage leads and temperature leads, respectively. Results are shown in Fig. \ref{fig:SuppFig-Seebeck}. The value near room temperature for our crystal with $x = 0$ is consistent with previous reports \cite{kuropatwa2012thermoelectric}. The smaller positive Seebeck coefficient seen for $x = 0.61$ is consistent with indium substitution in \sibt\ increasing the hole concentration.

\begin{figure*}
\begin{center}
\includegraphics[width=3.5in]{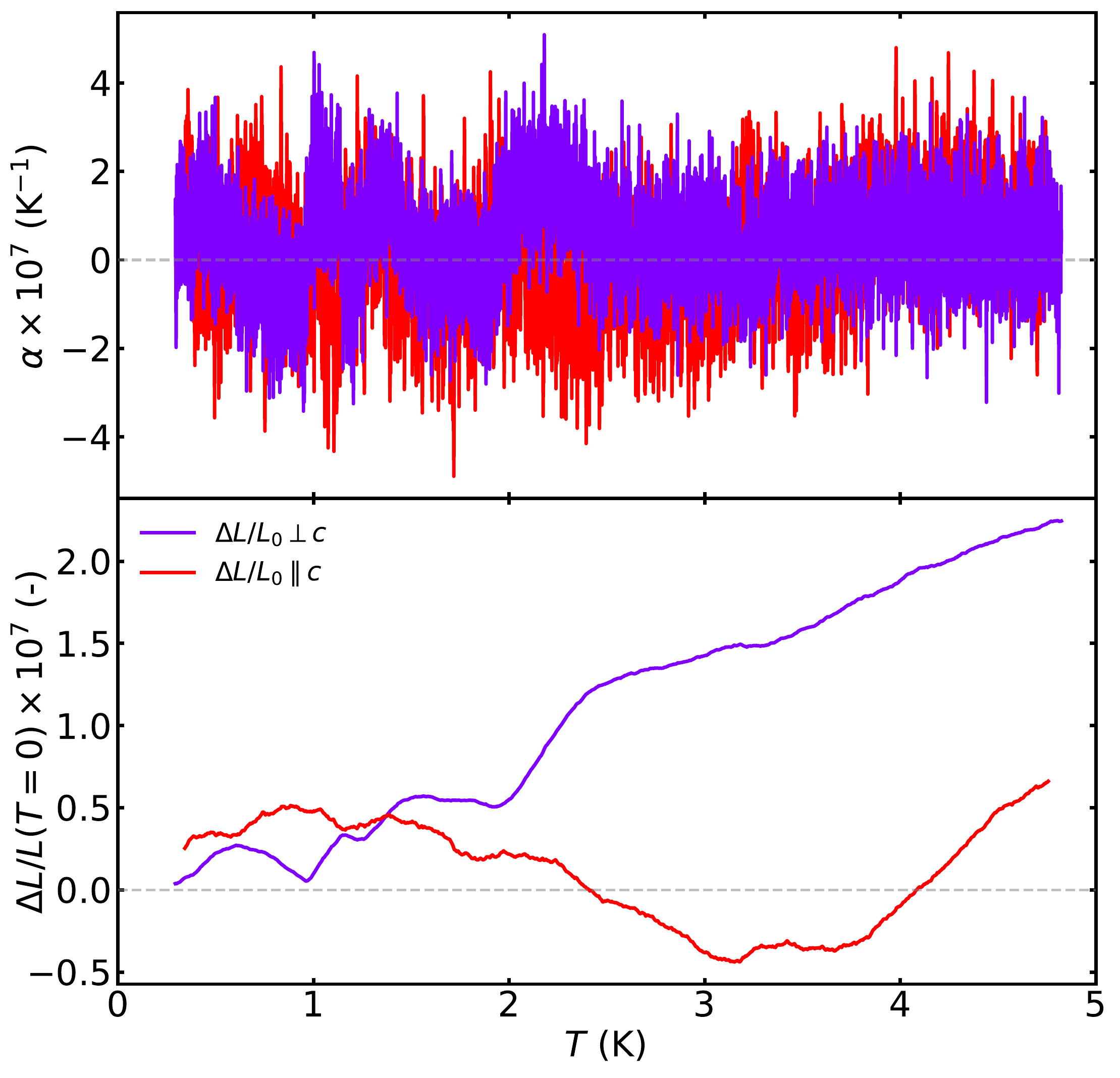}
\caption{\label{fig:SuppFig-thermal_expansion}
Results of thermal expansion measurements from a crystal with $x = 0.61$, showing no resolvable anomaly at $T_c$.
}
\end{center}
\end{figure*}

Linear thermal expansion of an \sibt\ crystal with $x = 0.61$ ($T_c$ = 1.85\,K) was measured in a CMR adiabatic demagnetization refrigerator between 0.2 and 5 K using a capactive dilatometer as described in Ref. \citenum{schmiedeshoff2006versatile}. Length changes $\Delta L$ were measured along and perpendicular to the crystallographic c-axis with lengths $L_0$ = 0.3 and 1 mm, respectively. The linear thermal expansion coefficient $\alpha = (1/L_0) d\Delta L/dT$ was computed after smoothing the temperature dependence of $\Delta L$. The results are shown in Fig. \ref{fig:SuppFig-thermal_expansion}.

\begin{figure*}
\begin{center}
\includegraphics[width=3.5in]{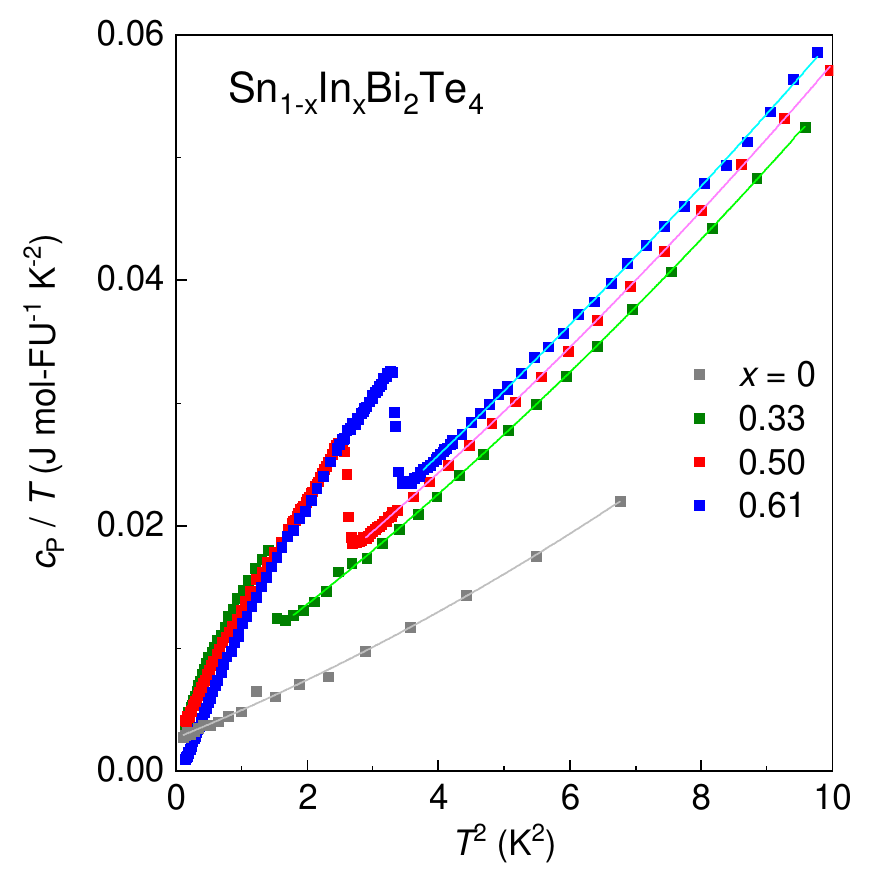}
\caption{\label{fig:SuppFig-hc}
Fits to normal state specific heat capacity data.
}
\end{center}
\end{figure*}

The low-temperature specific heat capacity data was analyzed as shown in Fig. \ref{fig:SuppFig-hc}. Data in the normal state is well described by  $c_P(T) = \gamma T + \beta T^3 + \delta T^5$, where the first term is the electronic contribution (with Sommerfeld coefficient, $\gamma$) and the second and third terms describe the phonon contribution. The data are plotted as $c_P/T$ vs $T^2$, and the slight curvature seen in the plots warranted inclusion of the $\delta T^5$ term. The parameters determined by the fits are collected in Table \ref{hcfits}.

\begin{figure*}
\begin{center}
\includegraphics[width=6.0in]{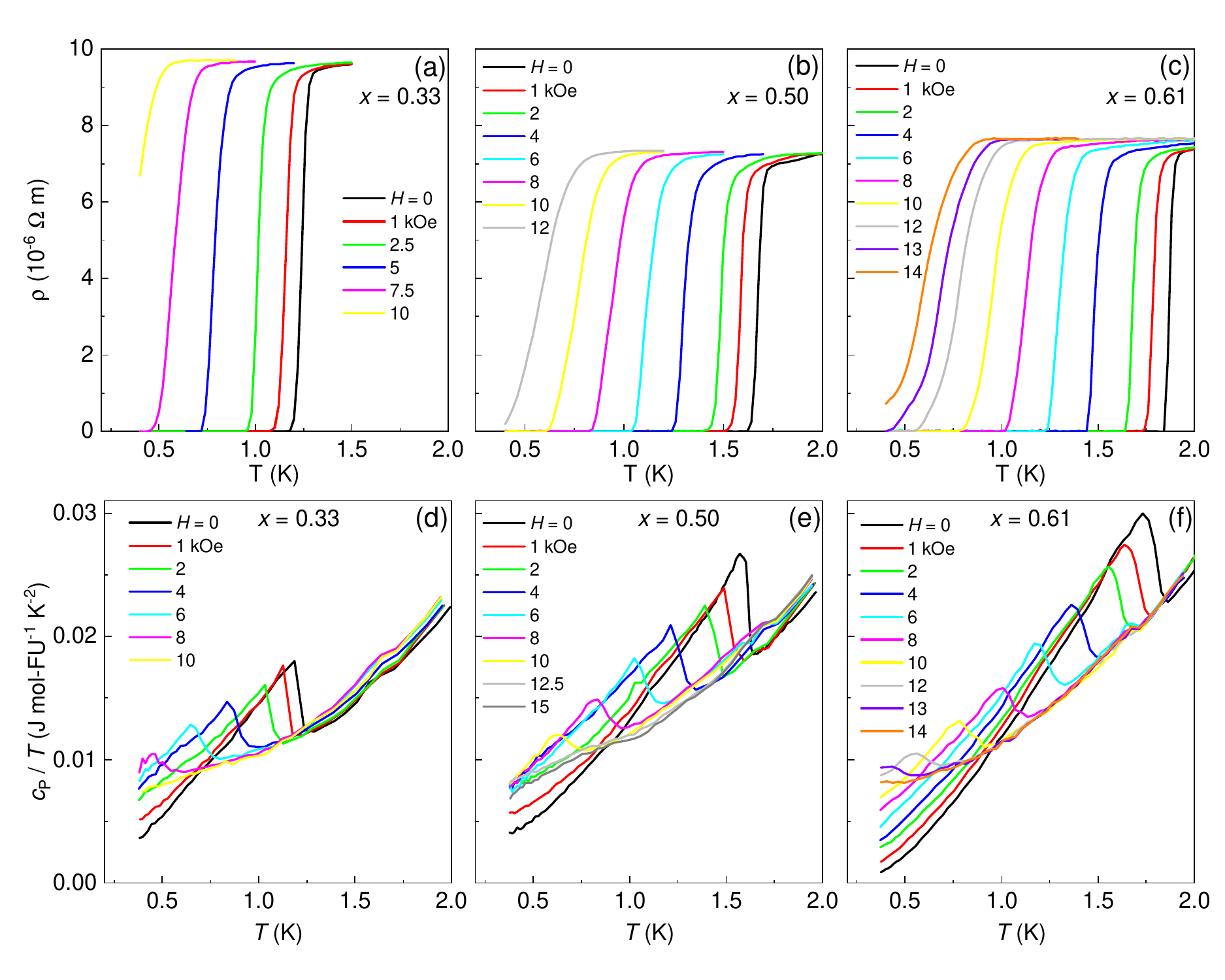}
\caption{\label{fig:SuppFig-rhocp}
Magnetic field dependence of superconducting transitions in \sibt\ seen in electrical resistivity (a-c) and specific heat capacity plotted as $c_P/T$ (d-f).
}
\end{center}
\end{figure*}

Additional heat capacity and electrical resistivity data used to construct the phase diagrams are shown in Fig. \ref{fig:SuppFig-rhocp}.

ARPES measurements were performed in a laboratory-based system using a Scienta DA30L electron spectrometer and an Oxide h$\nu$ = 11 eV laser light source.  Bulk crystals were mounted to a sample holder using silver epoxy with a cleaving post mounted on top.  The crystals were cleaved and measured at a base temperature of T$\sim$8 K and base pressure below $5 \times 10^{-11}$ Torr by knocking off the top post.  Linear polarization (p-wave) light was used, and the instrument was set for a total energy resolution of $\sim$4 meV and angular resolution of $\sim$0.01{\AA}$^{-1}$.

Neutron diffraction data from a rectangular platelet-like crystal of \sbt\ with the dimensions of $3.90\times2.98\times0.24$ mm$^3$  were measured on the TOPAZ single-crystal neutron diffractometer at the ORNL Spallation Neutron Source (SNS) \cite{coates2018}. The crystal was glued on the tip of a MiTeGen loop using Super Glue and transferred to the TOPAZ ambient goniometer for data collection at room temperature. The TOPAZ instrument uses the neutron wavelength-resolved Laue method for data collection by expanding the measured diffraction pattern from 2D on detector spaces (x, y) to wavelength‐resolved 3D volume in (x, y, $\lambda$) along the neutron time-of-flight direction. The sample orientations were optimized with the CrystalPlan software for better than 98\% coverage based on crystal symmetry \cite{zikovsky2011}. A total of 9 sample orientations were used in data collection. Each of the sample orientations was measured for approximately 1 h with 5 C of proton charge with SNS beam power at 1.4 MW. The integrated raw Bragg intensities were obtained using the 3-D ellipsoidal Q-space integration following previously reported methods \cite{schultz2014}. Data reduction, including neutron TOF spectrum, Lorentz, and detector efficiency corrections, was carried out with the mantid python program \cite{michels2016}. The reduced data were saved in SHELX HKLF2 wavelength-resolved Laue format, in which the wavelength is recorded separately for each reflection, and data were not merged. Starting with the reported structure refined from our powder x-ray diffraction data, the crystal structure was refined using the JANA2020 program \cite{petricek2014}. Refinement results are collected in Tables \ref{neutron} and \ref{atoms}.

\begin{table*}
\caption{Single crystal neutron diffraction results for \sbt.}
\begin{tabular}{lc}
\hline															
Chemical formula	&	Sn$_{0.986}$Bi$_{2.014}$Te$_4$	\\
Molar mass	&	1048.3	\\
Crystal system, space group	&	Trigonal, $R\overline{3}m$	\\
Temperature (K)	&	295	\\
a, c ({\AA})	&	4.4039 (6), 41.455 (8)	\\
V ({\AA}$^3$)	&	696.28 (19)	\\
Z	&	3	\\
Radiation type	&	Neutron, $\lambda$ = 0.4006-3.4989 {\AA}	\\
$\mu$ (mm$^{-1}$)	&	0.0353 + 0.00945$\lambda$	\\
Crystal size (mm)	&	3.90 $\times$ 2.98 $\times$ 0.24	\\
Diffractometer	&	TOPAZ	\\
T$_{min}$, T$_{max}$	&	0.947, 0.974	\\
R[$F^2 > 2\sigma(F^2)$], wR($F^2$), S	&	0.053, 0.132, 1.94	\\
No. of reflections	&	2717	\\
No. of parameters	&	34	\\
\hline															
\end{tabular}\
\label{neutron}
\end{table*}

\begin{table*}
\caption{Refined atomic coordinates, occupancies, and equivalent isotropic displacement parameters \sbt.}
\begin{tabular}{lccccc}
\hline															
atom	&	x	&	y	&	z	&	U$_{eq}$	&	occupancy	\\
\hline
Te1	&	0	&	0	&	0.135455(17)	&	0.01851(18)	&	1	\\
Te2	&	0	&	0	&	0.289102(13)	&	0.01665(14)	&	1	\\
Sn1	&	0	&	0	&	0	&	0.0230(3)	&	 0.903(15)	\\
Bi1	&	0	&	0	&	0	&	0.0230(3)	&	0.097(15)	\\
Bi2	&	0	&	0	&	0.428332(11)	&	0.02380(15)	&	0.959(12)	\\
Sn2	&	0	&	0	&	0.428332(11)	&	0.02380(15)	&	0.041(12)	\\

\hline															
\end{tabular}\
\label{atoms}
\end{table*}

\begin{table*}
\caption{Fit parameters from the normal state heat capacity of \sibt\ shown in the main text, and the Debye temperature $\Theta_D$ calculated from $\beta$. }
\begin{tabular}{lcccc}
\hline															
$x$ in 	&	$\gamma$ 	&	$\beta$	&	$\delta$	&	$\Theta_D$	\\
\sibt	&	(J mol-FU$^{-1}$ K$^{-2}$)	&	(J mol-FU$^{-1}$ K$^{-4}$)	&	(J mol-FU$^{-1}$ K$^{-6}$)	&	(K)	\\
\hline									
0	&	0.0027(1)	&	0.0022(1)	&	0.00010(2)	&	184	\\
0.33	&	0.0055(2)	&	0.0038(1)	&	0.00011(1)	&	153	\\
0.50	&	0.0065(1)	&	0.0040(5)	&	0.000111(5)	&	150	\\
0.61	&	0.0065(4)	&	0.0045(1)	&	0.00008(1)	&	144	\\
\hline															
\end{tabular}\
\label{hcfits}
\end{table*}

%
%


%

\end{document}